\documentclass[aps,prb,twocolumn,showpacs,groupedaddress]{revtex4-1}
\usepackage{graphicx}
\begin{document}

\title{Enhancing the Critical Current of a Superconducting Film in a Wide Range of Magnetic Fields with a Conformal Array of Nanoscale Holes}

\author{Y. L. Wang$^{1}$}\email{ylwang@anl.gov}
\author{M. L. Latimer$^{1,2}$}
\author{Z. L. Xiao$^{1,2}$}\email{xiao@anl.gov}
\author{R. Divan$^{3}$}
\author{L. E. Ocola$^{3}$}
\author{G. W. Crabtree$^{1,4}$}
\author{W. K. Kwok$^{1}$}

\affiliation{$^{1}$1 Materials Science Division, Argonne National Laboratory, Argonne, Illinois 60439, USA}

\affiliation{$^{2}$Department of Physics, Northern Illinois University, DeKalb, Illinois 60115, USA}

\affiliation{$^{3}$Center for Nanoscale Materials, Argonne National Laboratory, Argonne, Illinois 60439, USA}

\affiliation{$^{4}$Departments of Physics, Electrical and Mechanical Engineering, University of Illinois at Chicago, Illinois 60607, USA}

\date{\today}

\begin{abstract}
The maximum current (critical current) a type-II superconductor can transmit without energy loss is limited by the motion of the quantized magnetic flux penetrating into a superconductor. Introducing nanoscale holes into a superconducting film has been long pursued as a promising way to increase the critical current. So far the critical current enhancement was found to be mostly limited to low magnetic fields. Here we experimentally investigate the critical currents of superconducting films with a conformal array of nanoscale holes that have non-uniform density while preserving the local ordering. We find that the conformal array of nanoscle holes provides a more significant critical current enhancement at high magnetic fields. The better performance can be attributed to its arching effect that not only gives rise to the gradient in hole-density for pinning vortices with a wide range of densities but also prevent vortex channeling occurring in samples with a regular lattice of holes.
\end{abstract}


\maketitle

Critical current ($I_c$), below which a superconductor can transmit electrical power without energy loss, is a parameter of primary importance for potential applications of the material.\cite{Ref1} In a type-II superconductor, it is limited essentially by the motion of vortices, each consisting of exactly one quantum of flux ($\Phi_0= hc/2e = 20.7$ $G \cdot \mu m^2$) surrounded by circulating supercurrents in the plane perpendicular to the magnetic field.\cite{Ref2} In order to overcome this limitation, various types of artificial pinning centers are introduced into a superconductor to immobilize the vortices.\cite{Ref1,Ref2,Ref3,Ref4,Ref5,Ref6,Ref7,Ref8,Ref9,Ref10,Ref11,Ref12,Ref13,Ref14,Ref15,Ref16,Ref17,Ref18,Ref19}  Among them, regular arrays of holes in superconducting films\cite{Ref9,Ref10,Ref11,Ref12,Ref13,Ref14,Ref15,Ref16,Ref17,Ref18,Ref19,New20,New21} have been explored to enhance vortex pinning and hence the current-carrying capacity of a superconductor. It has been shown that an enhancement in the critical current can be achieved when the vortex lattice is commensurate with the underlying periodic hole array.\cite{Ref12,Ref13,Ref14,Ref15,Ref16} That is, local maxima appear in the magnetic field dependence of the critical current $I_c(H)$ at matching fields at which the density of vortices equals an integer multiple of that of the holes.  Away from the matching fields, however, the enhancement of the critical current is reduced. In order to overcome this shortcoming, more complicated pinning topologies such as quasiperiodic arrays were proposed.\cite{Ref20,Ref21,Ref22} Due to the existence of multiple periodicities in these arrays, more maxima or extended peaks are expected to occur and have been experimentally observed in Nb films\cite{Ref14,Ref22} and Pb film\cite{Ref23,Ref24} containing quasiperiodic Penrose lattices of artificial pinning centers.

Theoretical and experimental studies indicate that graded pinning landscapes can be an excellent candidate for controlling vortex motion.\cite{Ref25,Ref26,Ref27,Ref28} By changing the hole density in the equally spaced rows parallel to the current flow, Wu et al. observed a ratchet effect in patterned Nb films.\cite{Ref25} Motta et al. conducted magnetization measurements on a MoGe film with rows of equal density of holes but with inwardly increased row separation and found suppressed avalanche and increased critical currents in comparison to films with a square array of holes.\cite{Ref26} Misko and Nori proposed to place pinning sites on vertices of hyperbolic tessellations to trap vortices in a broad range of the applied magnetic field and to serve as a ‘capacitor’ to store vortices.\cite{Ref27}

 \begin{figure}
 \includegraphics[width=0.42\textwidth]{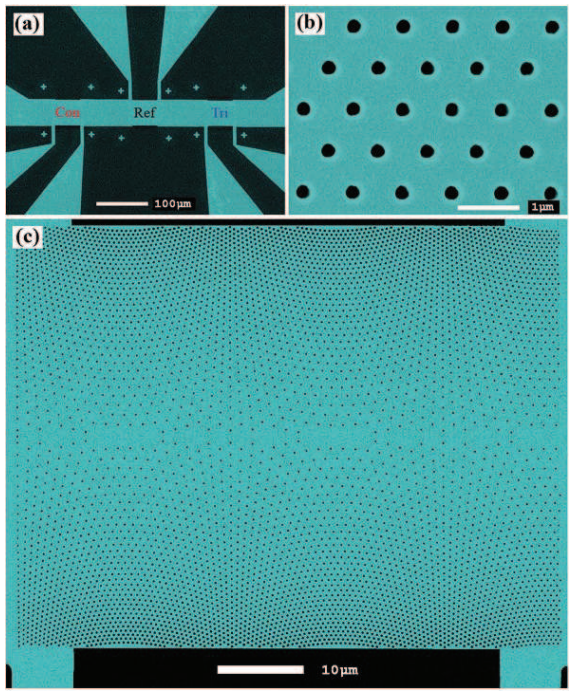}%
 \caption{\label{fig:fig1}(Color online) \textbf{SEM micrographs of Sample I.} (a),arrangement of three sections. (b) and (c) are images at higher magnifications for the triangular and conformal arrays, respectively.}
 \end{figure}
Conformal crystals are topologically perfect, two-dimensional (2D) structures created through a conformal (angle-preserving) transfermation to a regular lattice(Fig. \ref{fig:fig1}c).\cite{Ref28} This notion was introduced by Rothen et. al.\cite{Ref29,Ref30} to describe the ‘gravity rainbow’ distribution of magnetic spheres subjected to an external force such as gravity and/or magnetic field gradient. Conformal arrangement has been widely used in, for example, designing optic devices,\cite{Ref31} engineering antenna patterns,\cite{Ref32}  confining ultrasound,\cite{Ref33} and waveguiding.\cite{Ref33,Ref34}  In recent computer simulations on the magnetization of patterned superconducting films, Ray et. al. demonstrated  that a conformal pinning array transformed of a uniform triangular lattice produces a much higher critical current over a wider range of magnetic fields than any pinning geometry considered up until now.\cite{Ref28} They attributed the pinning enhancement to the unique arrangement of the pinning sites in the conformal crystals which not only have a density gradient, but also preserve aspects of the local sixfold symmetry naturally adopted by a vortex lattice.

Here we report transport measurements on the critical currents of superconducting films containing direct and conformal arrays of triangular hole-lattices as well as the unpatterned ones to reveal the advantages of conformal pinning arrays on the current carrying capacity. We observed higher critical currents over a large range of the applied magnetic field in films with a conformal lattice of holes compared to those with a triangular array of holes. 

Experiments were carried out on MoGe superconducting thin films which are known for their weak random pinning,\cite{Ref35} enabling transport measurements at temperatures far away from the  zero-field critical temperature $T_{c0}$ to avoid complications from phenomena originated from the Little-Parks effect which can appear in patterned films near $T_{c0}$.\cite{Ref36,New39,New40}  In order to directly compare the effects of the various pinning arrays on the current carrying capacity, we placed three equivalent voltage pairs at various sections of the same microbridge (Fig.\ref{fig:fig1}a). One section of the microbridge was kept unpatterned as a reference (Ref). Triangular (Tri) (Fig.\ref{fig:fig1}b) and conformal (Con) (Fig.\ref{fig:fig1}c) arrays of holes, which have equivalent average density of nanoscle holes, were introduced into the other two sections through electron-beam lithography, followed by reactive ion etching. Transport measurements were carried out using a standard dc four-probe method. The current flows horizontally along the long length of the microbridge. The voltages of the three bridges in each sample were measured at the same time. The applied magnetic field is always perpendicular to the film plane. The detailed measurements and sample fabrications including creating conformal lattice arrays can be found in Ref\cite{Sup}.

\begin{figure}
 \includegraphics[width=0.45\textwidth]{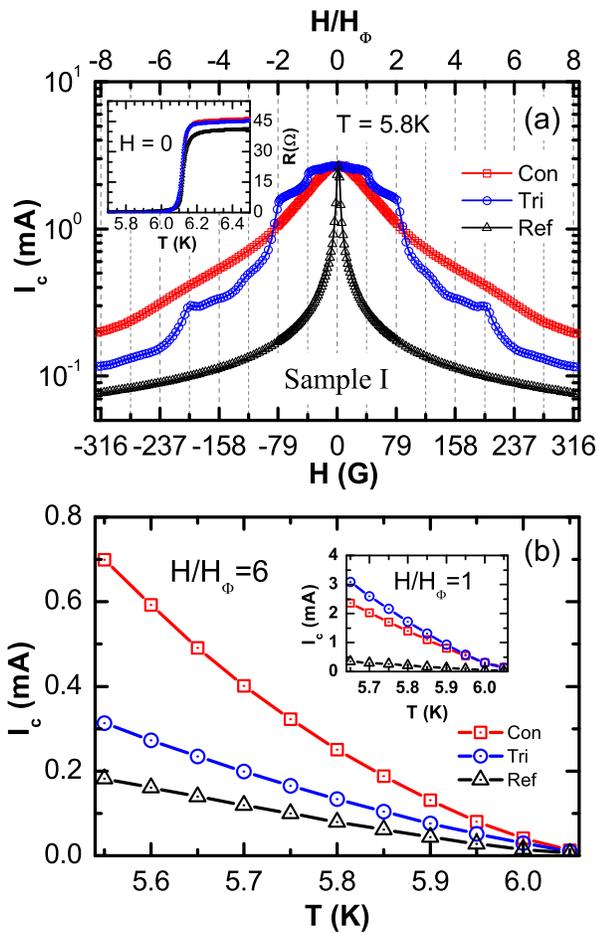}%
 \caption{\label{fig:fig2}(Color online) \textbf{Critical currents from Sample I.} The magnetic field and temperature dependences of critical currents are presented in (a) and (b), respectively. The inset in (a) shows the zero-field superconducting transitions of the three sections.}
 \end{figure}

Three samples with various lattice constants, hole diameters and hole depth for the hole-arrays were investigated and their parameters are given in Table \ref{tab:tab1}.
 \begin{table}
 \caption{\label{tab:tab1}\textbf{Sample parameters.} $d$, diameter of the etched holes;
$b_1$, lattice constant of the direct triangular array from which the conformal array is transformed;
$b_2$, hole-hole spacing of the triangular array with the equivalent average hole density as that of its conformal counterpart;
$H_\Phi$, matching field;
$T_(c0)$, zero-field superconducting critical temperature.}
 \begin{ruledtabular}
 \begin{tabular}{c|c|c|c|c|c|c}
                                 &                                 & $b_1$ & $b_2$ & $H_\Phi$ &                                    & hole\\
                                 & \raisebox{1.5ex}[0pt]{$d$ (nm)} & (nm)  & (nm)  & (G)      & \raisebox{1.5ex}[0pt]{$T_(c0)$ (K)}   & type\\
  \hline
 I & $220\pm 19$ & $500$ & $777$ & $39.5$ & $6.152\pm 0.004$ & through\\
II & $110\pm 10$ & $300$ & $466$ & $110$ & $6.163\pm 0.002$  & through\\
III & $110\pm 10$ & $300$ & $466$ & $110$ & $6.069\pm 0.007$ & blind\\
 \end{tabular}
 \end{ruledtabular}
 \end{table}
The inset of Fig.\ref{fig:fig2}a shows the temperature dependence of the resistances for the three sections of sample I at zero magnetic field. The normal state resistances of the patterned sections are slightly larger than that of the pristine reference one, due to the presence of a hole structure which reduces the effective area of the cross-section for the current flow. The superconducting transition of all three sections remains the same, indicating no degradation in their film quality arising from the patterning process and enabling a direct comparison of the pinning effect between the triangular and conformal hole-arrays.

The main panel of Fig.\ref{fig:fig2}a presents the magnetic field dependence of the critical current $I_c(H)$ for all three sections of sample I at the same temperature ($T = 5.8$ K). Evidently both patterned sections have higher critical currents than those of the unpatterned reference section at all applied magnetic fields. Since the values of the critical currents for all sections at zero magnetic field are the same, their enhancement in the patterned sections in external magnetic fields should come from the vortex pinning effect of the fabricated hole-arrays.  At the calculated first ($H = H_\Phi$) and second ($H = 2H_\Phi$) matching fields we observed steps instead of narrow peaks in the $I_c(H)$ curve for the section containing triangular lattice of holes. This is an indication of strong pinning of the holes, which overshadows the effects on the critical currents gained from the reduction of  the vortex interaction energy at matching, similar to that observed in perforated Nb films at low temperatures.\cite{Ref11} The steep drop in the critical current once the magnetic field exceeds the second matching value indicates the appearance of interstitial vortices which can be depinned at a much lower driving force in a regular pinning array with easy vortex flow channels.\cite{Ref28,Ref37} That is, at low magnetic fields only two vortices are allowed to sit in each hole, consistent with the theoretical saturation number $n_{si} = d/4\xi(T) \approx 2$ for an isolated hole, where $\xi(T) = \xi(0)(1-T/T_{c0})^{-1/2}$ is the temperature dependent superconducting coherence length, using $\xi(0) = 6$ nm derived from the critical temperature versus magnetic field phase diagram\cite{Ref38} and $d = 220$ nm for the diameter of the holes in Sample I. Above the second matching field, the depinning of the weakly pinned interstitial vortices makes the third and fourth matchings almost indistinguishable. On the other hand, a small peak appears at the fifth matching field, possibly due to the caging effect similar to that observed in MoGe films containing honeycomb hole-arrays.\cite{Ref38}

The $I_c(H)$ curve (red) in Fig.\ref{fig:fig2}a for the section with a conformal hole-array shows a smooth decay in the critical current with increasing magnetic field. Since the holes in all sections are fabricated under the same conditions, each hole in the conformal lattice should also be able to hold at least two vortices.  However, there exists no noticeable feature (e.g. steps or bumps) at the first and second matching fields where the density of the vortices is respectively equal to and twice of that of the average hole-density, in contrast to the presence of obvious steps in the $I_c(H)$ curve for the section with a triangular hole-array. This is due to the hole-density gradient which decreases gradually from the edge towards the middle of bridge (Fig. \ref{fig:fig1}c). That is, the vortex lattice is commensurate with the hole-arrangement and lead to the appearance of the matching effect locally. However, due to the gradient in the hole-hole separation, there can exist various local matching fields, smearing out the bulk matching field features, e.g. steps in the $I_c(H)$ curve. Although the multiple periods of conformal array can result in pinning enhancement over a wide range of magnetic field (rather than a peak at one matching field),\cite{Ref14,Ref20,Ref21,Ref22} the incommensurate conformal array with the Abrikosov triangular lattice will lead to an increase in the interaction energy in the vortex matter, thus reducing the critical current. This is in fact consistent with our data in Fig.\ref{fig:fig2}a: the triangular hole-array which is perfectly commensurate with the Abrikosov triangular lattice can lead to better vortex pinning when all vortices are pinned inside the holes at magnetic fields lower than $2H_\Phi$.  On the other hand, the conformal hole-array outperforms its triangular array counterpart in pinning efficiency once interstitial vortices set in beyond $2H_\Phi$: contrasted with the triangular hole-array, the critical current for the section with conformal hole-array is devoid of steps in the critical current and quickly surpasses that of the section with a triangular hole-lattice above $2H_\Phi$.

Figure \ref{fig:fig2}b presents the temperature dependence of the critical current $I_c(T)$ for applied magnetic fields at the first (inset) and sixth (main panel) matching fields, respectively. At all experimental temperatures the patterned sections have higher critical currents than those of the unpatterned one, and the conformal array outperforms the triangular array at high magnetic fields ($H/H_\Phi = 6$) while the latter does better at low fields ($H/H_\Phi = 1$). Since interstitial vortices appear at $H > 2H_\Phi$ in the section with the triangular hole-array, the larger critical current enhancement induced by the conformal hole-array indicates that it is more effective than the triangular hole-array in pinning interstitial vortices. This is consistent with that observed in computer simulations: the curved hole-array geometry in a conformal array prevents the formation of easy channels of vortex flow.\cite{Ref28} 

In order to further confirm that a conformal array indeed has advantages in pinning interstitial vortices, we fabricated a sample (Sample II) with smaller holes to reduce the vortex saturation number $n_{si}$ to $1$. We wanted to see whether the critical currents of a conformal hole-array section would surpass that of the triangular hole-lattice at magnetic fields above $H_\Phi$ instead of $2H_\Phi$ observed in Sample I.
 \begin{figure}
 \includegraphics[width=0.45\textwidth]{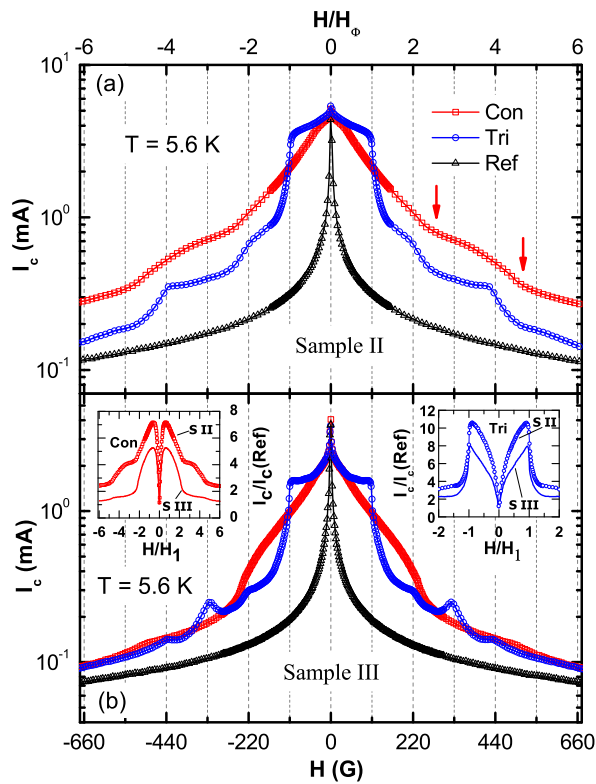}%
 \caption{\label{fig:fig3}(Color online) \textbf{Magnetic field dependence of critical currents from Sample II and Sample III.} Sample II (a) and sample III (b) has half the size of holes as those in Sample I. Sample III has the same arrangement and diameter but half the depth of the holes as those in Sample II. The insets in Fig.\ref{fig:fig3}b show the difference in the enhancement of the critical currents induced by through and blind hole-arrays in Samples II and III, the left and right insets are for the conformal and triangular arrays, respectively.}
 \end{figure}
Data for Sample II obtained at $T = 5.6$ K are presented in Fig.\ref{fig:fig3}a. As expected, the critical current of the section containing a triangular hole-array drops sharply at the first matching field $H_\Phi$, revealing a saturation number of $1$.  At $H  > H_\Phi$, the conformal hole-array evidently performs better than its triangular counterpart.  It is also worthy noting that there exist specific magnetic field values (indicated by arrows) at which the rate of decrease of the critical current in the section with conformal hole-array is somewhat abated. The field value at the first arrow is very close to the first matching field of $265$ G calculated for the denser hole-array near the edge of the microbridge, which can be approximated as a triangular array with a lattice constant of $300$ nm. That is, the conformal hole-array becomes more effective in pinning when all holes are occupied by vortices.  On the other hand, our data also indicate that interstitial vortices start to appear in the conformal array, e.g. in the middle area of the microbridge, long before all the holes are filled with vortices. The field value at the second arrow is roughly double that of the first one, probably due to the local commensurate effect between the vortex lattice and the hole-array near the edge area.

In the computer simulations by Ray et al., the advantages of a conformal array gradually varnishes with decreasing strength of the pinning sites.\cite{Ref28} In experiments we can fabricate holes with different pinning strengths by varying their depth. This can be realized through controlling the time of reactive ion etching. In Samples I and II, we used a two-minute etch time to create through holes.  For Sample III, we reduced the etching time to one minute to form blind holes with a depth roughly half of the film thickness. Sample III is simply a duplicate of Sample II, except for the difference in the depth of the holes.

Figure \ref{fig:fig3}b shows the critical currents of Sample III. The effect of the hole depth on the vortex pinning strengths can be shown by comparing the critical currents of the patterned sections to the reference one on the same sample. The left and right insets in Fig.\ref{fig:fig3}b present these comparisons for the conformal and triangular arrays, respectively, indicating weaker pinning strength of the holes in Sample III. Consistent with that observed by Ray et. al.,\cite{Ref28} the conformal hole-array in sample III has no significant advantages over the triangular lattice which can even pin interstitial vortices better and have higher critical currents due to caging effect \cite{New44} (e.g. at and near the third matching field). This is understandable that the interaction energy due to the deformation of the vortex lattice can become dominated when the pinning strength in a conformal hole-array is weakened. With a strong pinning strength, our experimental setup in which the applied current flows along the hole-row (see Fig.\ref{fig:fig1}b) is the best scenario in blocking the motion of interstitial vortices in a triangular hole-lattice. That said, the advantages of conformal array in enhancing the critical current would be even more clearly seen if the current were applied in a direction which enables the interstitial vortices to move parallel to the hole-row in the triangular lattice or parallel to one of the principal axis in a square lattice.

In conclusion, we investigated the pinning effect of a conformal hole-array by comparing it with a triangular counterpart and that without holes. Although both the conformal and triangular arrays can enhance the critical current at all applied magnetic fields, the latter can pin vortices more effectively when they sit in the holes.  However, a conformal lattice of holes with strong pinning strength has significant advantages in preventing the motion of interstitial vortices, enhancing the critical currents at high magnetic fields. On the other hand, a triangular lattice can have advantages in pinning interstitial vortices due to caging effect when the strength of the pinning sites is weak.

\begin{acknowledgments} This work was supported by DOE BES under Contract No. DE-AC02-06CH11357 which also funds Argonne’s Center for Nanoscale Materials (CNM) and Electron Microscopy Center (EMC) where the nanopatterning and morphological analysis were performed. M.L.L. and Z.L.X. acknowledge DOE BES Grant No. DE-FG02-06ER46334 (sample fabrication). We are grateful to Dr. Charles Reichhardt in Los Alamos National Laboratory for stimulating discussions and sharing their simulation results prior to publication.
\end{acknowledgments}

\bibliography{ConformalPrbM2}

\begin{thebibliography}{44}%
\makeatletter
\providecommand \@ifxundefined [1]{%
 \@ifx{#1\undefined}
}%
\providecommand \@ifnum [1]{%
 \ifnum #1\expandafter \@firstoftwo
 \else \expandafter \@secondoftwo
 \fi
}%
\providecommand \@ifx [1]{%
 \ifx #1\expandafter \@firstoftwo
 \else \expandafter \@secondoftwo
 \fi
}%
\providecommand \natexlab [1]{#1}%
\providecommand \enquote  [1]{``#1''}%
\providecommand \bibnamefont  [1]{#1}%
\providecommand \bibfnamefont [1]{#1}%
\providecommand \citenamefont [1]{#1}%
\providecommand \href@noop [0]{\@secondoftwo}%
\providecommand \href [0]{\begingroup \@sanitize@url \@href}%
\providecommand \@href[1]{\@@startlink{#1}\@@href}%
\providecommand \@@href[1]{\endgroup#1\@@endlink}%
\providecommand \@sanitize@url [0]{\catcode `\\12\catcode `\$12\catcode
  `\&12\catcode `\#12\catcode `\^12\catcode `\_12\catcode `\%12\relax}%
\providecommand \@@startlink[1]{}%
\providecommand \@@endlink[0]{}%
\providecommand \url  [0]{\begingroup\@sanitize@url \@url }%
\providecommand \@url [1]{\endgroup\@href {#1}{\urlprefix }}%
\providecommand \urlprefix  [0]{URL }%
\providecommand \Eprint [0]{\href }%
\providecommand \doibase [0]{http://dx.doi.org/}%
\providecommand \selectlanguage [0]{\@gobble}%
\providecommand \bibinfo  [0]{\@secondoftwo}%
\providecommand \bibfield  [0]{\@secondoftwo}%
\providecommand \translation [1]{[#1]}%
\providecommand \BibitemOpen [0]{}%
\providecommand \bibitemStop [0]{}%
\providecommand \bibitemNoStop [0]{.\EOS\space}%
\providecommand \EOS [0]{\spacefactor3000\relax}%
\providecommand \BibitemShut  [1]{\csname bibitem#1\endcsname}%
\let\auto@bib@innerbib\@empty
\bibitem [{\citenamefont {Chong}\ \emph {et~al.}(1997)\citenamefont {Chong},
  \citenamefont {Hiroi}, \citenamefont {Izumi}, \citenamefont {Shimoyama},
  \citenamefont {Nakayama}, \citenamefont {Kishio}, \citenamefont {Terashima},
  \citenamefont {Bando},\ and\ \citenamefont {Takano}}]{Ref1}%
  \BibitemOpen
  \bibfield  {author} {\bibinfo {author} {\bibfnamefont {I.}~\bibnamefont
  {Chong}}, \bibinfo {author} {\bibfnamefont {Z.}~\bibnamefont {Hiroi}},
  \bibinfo {author} {\bibfnamefont {M.}~\bibnamefont {Izumi}}, \bibinfo
  {author} {\bibfnamefont {J.}~\bibnamefont {Shimoyama}}, \bibinfo {author}
  {\bibfnamefont {Y.}~\bibnamefont {Nakayama}}, \bibinfo {author}
  {\bibfnamefont {K.}~\bibnamefont {Kishio}}, \bibinfo {author} {\bibfnamefont
  {T.}~\bibnamefont {Terashima}}, \bibinfo {author} {\bibfnamefont
  {Y.}~\bibnamefont {Bando}}, \ and\ \bibinfo {author} {\bibfnamefont
  {M.}~\bibnamefont {Takano}},\ }\href {\doibase 10.1126/science.276.5313.770}
  {\bibfield  {journal} {\bibinfo  {journal} {Science}\ }\textbf {\bibinfo
  {volume} {276}},\ \bibinfo {pages} {770} (\bibinfo {year}
  {1997})}\BibitemShut {NoStop}%
\bibitem [{\citenamefont {Crabtree}\ and\ \citenamefont {Nelson}(1997)}]{Ref2}%
  \BibitemOpen
  \bibfield  {author} {\bibinfo {author} {\bibfnamefont {G.~W.}\ \bibnamefont
  {Crabtree}}\ and\ \bibinfo {author} {\bibfnamefont {D.~R.}\ \bibnamefont
  {Nelson}},\ }\href {\doibase 10.1063/1.881715} {\bibfield  {journal}
  {\bibinfo  {journal} {Physics Today}\ }\textbf {\bibinfo {volume} {50}},\
  \bibinfo {pages} {38} (\bibinfo {year} {1997})}\BibitemShut {NoStop}%
\bibitem [{\citenamefont {Martin}\ \emph {et~al.}(1997)\citenamefont {Martin},
  \citenamefont {Velez}, \citenamefont {Nogues},\ and\ \citenamefont
  {Schuller}}]{Ref3}%
  \BibitemOpen
  \bibfield  {author} {\bibinfo {author} {\bibfnamefont {J.~I.}\ \bibnamefont
  {Martin}}, \bibinfo {author} {\bibfnamefont {M.}~\bibnamefont {Velez}},
  \bibinfo {author} {\bibfnamefont {J.}~\bibnamefont {Nogues}}, \ and\ \bibinfo
  {author} {\bibfnamefont {I.~K.}\ \bibnamefont {Schuller}},\ }\href {\doibase
  10.1103/PhysRevLett.79.1929} {\bibfield  {journal} {\bibinfo  {journal}
  {Phys. Rev. Lett.}\ }\textbf {\bibinfo {volume} {79}},\ \bibinfo {pages}
  {1929} (\bibinfo {year} {1997})}\BibitemShut {NoStop}%
\bibitem [{\citenamefont {Haugan}\ \emph {et~al.}(2004)\citenamefont {Haugan},
  \citenamefont {Barnes}, \citenamefont {Wheeler}, \citenamefont
  {Meisenkothen},\ and\ \citenamefont {Sumption}}]{Ref4}%
  \BibitemOpen
  \bibfield  {author} {\bibinfo {author} {\bibfnamefont {T.}~\bibnamefont
  {Haugan}}, \bibinfo {author} {\bibfnamefont {P.~N.}\ \bibnamefont {Barnes}},
  \bibinfo {author} {\bibfnamefont {R.}~\bibnamefont {Wheeler}}, \bibinfo
  {author} {\bibfnamefont {F.}~\bibnamefont {Meisenkothen}}, \ and\ \bibinfo
  {author} {\bibfnamefont {M.}~\bibnamefont {Sumption}},\ }\href@noop {}
  {\bibfield  {journal} {\bibinfo  {journal} {Nature}\ }\textbf {\bibinfo
  {volume} {430}},\ \bibinfo {pages} {867} (\bibinfo {year}
  {2004})}\BibitemShut {NoStop}%
\bibitem [{\citenamefont {Kang}\ \emph {et~al.}(2006)\citenamefont {Kang},
  \citenamefont {Goyal}, \citenamefont {Li}, \citenamefont {Gapud},
  \citenamefont {Martin}, \citenamefont {Heatherly}, \citenamefont {Thompson},
  \citenamefont {Christen}, \citenamefont {List}, \citenamefont {Paranthaman},\
  and\ \citenamefont {Lee}}]{Ref5}%
  \BibitemOpen
  \bibfield  {author} {\bibinfo {author} {\bibfnamefont {S.}~\bibnamefont
  {Kang}}, \bibinfo {author} {\bibfnamefont {A.}~\bibnamefont {Goyal}},
  \bibinfo {author} {\bibfnamefont {J.}~\bibnamefont {Li}}, \bibinfo {author}
  {\bibfnamefont {A.~A.}\ \bibnamefont {Gapud}}, \bibinfo {author}
  {\bibfnamefont {P.~M.}\ \bibnamefont {Martin}}, \bibinfo {author}
  {\bibfnamefont {L.}~\bibnamefont {Heatherly}}, \bibinfo {author}
  {\bibfnamefont {J.~R.}\ \bibnamefont {Thompson}}, \bibinfo {author}
  {\bibfnamefont {D.~K.}\ \bibnamefont {Christen}}, \bibinfo {author}
  {\bibfnamefont {F.~A.}\ \bibnamefont {List}}, \bibinfo {author}
  {\bibfnamefont {M.}~\bibnamefont {Paranthaman}}, \ and\ \bibinfo {author}
  {\bibfnamefont {D.~F.}\ \bibnamefont {Lee}},\ }\href {\doibase
  10.1126/science.1124872} {\bibfield  {journal} {\bibinfo  {journal}
  {Science}\ }\textbf {\bibinfo {volume} {311}},\ \bibinfo {pages} {1911}
  (\bibinfo {year} {2006})}\BibitemShut {NoStop}%
\bibitem [{\citenamefont {Llordes}\ \emph {et~al.}(2012)\citenamefont
  {Llordes}, \citenamefont {Palau}, \citenamefont {Gazquez}, \citenamefont
  {Coll}, \citenamefont {Vlad}, \citenamefont {Pomar}, \citenamefont {Arbiol},
  \citenamefont {Guzman}, \citenamefont {Ye}, \citenamefont {Rouco},
  \citenamefont {Sandiumenge}, \citenamefont {Ricart}, \citenamefont {Puig},
  \citenamefont {Varela}, \citenamefont {Chateigner}, \citenamefont {Vanacken},
  \citenamefont {Gutierrez}, \citenamefont {Moshchalkov}, \citenamefont
  {Deutscher}, \citenamefont {Magen},\ and\ \citenamefont {Obradors}}]{Ref6}%
  \BibitemOpen
  \bibfield  {author} {\bibinfo {author} {\bibfnamefont {A.}~\bibnamefont
  {Llordes}}, \bibinfo {author} {\bibfnamefont {A.}~\bibnamefont {Palau}},
  \bibinfo {author} {\bibfnamefont {J.}~\bibnamefont {Gazquez}}, \bibinfo
  {author} {\bibfnamefont {M.}~\bibnamefont {Coll}}, \bibinfo {author}
  {\bibfnamefont {R.}~\bibnamefont {Vlad}}, \bibinfo {author} {\bibfnamefont
  {A.}~\bibnamefont {Pomar}}, \bibinfo {author} {\bibfnamefont
  {J.}~\bibnamefont {Arbiol}}, \bibinfo {author} {\bibfnamefont
  {R.}~\bibnamefont {Guzman}}, \bibinfo {author} {\bibfnamefont
  {S.}~\bibnamefont {Ye}}, \bibinfo {author} {\bibfnamefont {V.}~\bibnamefont
  {Rouco}}, \bibinfo {author} {\bibfnamefont {F.}~\bibnamefont {Sandiumenge}},
  \bibinfo {author} {\bibfnamefont {S.}~\bibnamefont {Ricart}}, \bibinfo
  {author} {\bibfnamefont {T.}~\bibnamefont {Puig}}, \bibinfo {author}
  {\bibfnamefont {M.}~\bibnamefont {Varela}}, \bibinfo {author} {\bibfnamefont
  {D.}~\bibnamefont {Chateigner}}, \bibinfo {author} {\bibfnamefont
  {J.}~\bibnamefont {Vanacken}}, \bibinfo {author} {\bibfnamefont
  {J.}~\bibnamefont {Gutierrez}}, \bibinfo {author} {\bibfnamefont
  {V.}~\bibnamefont {Moshchalkov}}, \bibinfo {author} {\bibfnamefont
  {G.}~\bibnamefont {Deutscher}}, \bibinfo {author} {\bibfnamefont
  {C.}~\bibnamefont {Magen}}, \ and\ \bibinfo {author} {\bibfnamefont
  {X.}~\bibnamefont {Obradors}},\ }\href@noop {} {\bibfield  {journal}
  {\bibinfo  {journal} {Nat Mater}\ }\textbf {\bibinfo {volume} {11}},\
  \bibinfo {pages} {329} (\bibinfo {year} {2012})}\BibitemShut {NoStop}%
\bibitem [{\citenamefont {Hua}\ \emph {et~al.}(2010)\citenamefont {Hua},
  \citenamefont {Welp}, \citenamefont {Schlueter}, \citenamefont {Kayani},
  \citenamefont {Xiao}, \citenamefont {Crabtree},\ and\ \citenamefont
  {Kwok}}]{Ref7}%
  \BibitemOpen
  \bibfield  {author} {\bibinfo {author} {\bibfnamefont {J.}~\bibnamefont
  {Hua}}, \bibinfo {author} {\bibfnamefont {U.}~\bibnamefont {Welp}}, \bibinfo
  {author} {\bibfnamefont {J.}~\bibnamefont {Schlueter}}, \bibinfo {author}
  {\bibfnamefont {A.}~\bibnamefont {Kayani}}, \bibinfo {author} {\bibfnamefont
  {Z.~L.}\ \bibnamefont {Xiao}}, \bibinfo {author} {\bibfnamefont {G.~W.}\
  \bibnamefont {Crabtree}}, \ and\ \bibinfo {author} {\bibfnamefont {W.~K.}\
  \bibnamefont {Kwok}},\ }\href {\doibase 10.1103/PhysRevB.82.024505}
  {\bibfield  {journal} {\bibinfo  {journal} {Phys. Rev. B}\ }\textbf {\bibinfo
  {volume} {82}},\ \bibinfo {pages} {024505} (\bibinfo {year}
  {2010})}\BibitemShut {NoStop}%
\bibitem [{\citenamefont {Matsui}\ \emph {et~al.}(2012)\citenamefont {Matsui},
  \citenamefont {Ogiso}, \citenamefont {Yamasaki}, \citenamefont {Kumagai},
  \citenamefont {Sohma}, \citenamefont {Yamaguchi},\ and\ \citenamefont
  {Manabe}}]{Ref8}%
  \BibitemOpen
  \bibfield  {author} {\bibinfo {author} {\bibfnamefont {H.}~\bibnamefont
  {Matsui}}, \bibinfo {author} {\bibfnamefont {H.}~\bibnamefont {Ogiso}},
  \bibinfo {author} {\bibfnamefont {H.}~\bibnamefont {Yamasaki}}, \bibinfo
  {author} {\bibfnamefont {T.}~\bibnamefont {Kumagai}}, \bibinfo {author}
  {\bibfnamefont {M.}~\bibnamefont {Sohma}}, \bibinfo {author} {\bibfnamefont
  {I.}~\bibnamefont {Yamaguchi}}, \ and\ \bibinfo {author} {\bibfnamefont
  {T.}~\bibnamefont {Manabe}},\ }\href {\doibase 10.1063/1.4769836} {\bibfield
  {journal} {\bibinfo  {journal} {Applied Physics Letters}\ }\textbf {\bibinfo
  {volume} {101}},\ \bibinfo {eid} {232601} (\bibinfo {year}
  {2012})}\BibitemShut {NoStop}%
\bibitem [{\citenamefont {Velez}\ \emph {et~al.}(2008)\citenamefont {Velez},
  \citenamefont {Martin}, \citenamefont {Villegas}, \citenamefont {Hoffmann},
  \citenamefont {Gonzalez}, \citenamefont {Vicent},\ and\ \citenamefont
  {Schuller}}]{Ref9}%
  \BibitemOpen
  \bibfield  {author} {\bibinfo {author} {\bibfnamefont {M.}~\bibnamefont
  {Velez}}, \bibinfo {author} {\bibfnamefont {J.}~\bibnamefont {Martin}},
  \bibinfo {author} {\bibfnamefont {J.}~\bibnamefont {Villegas}}, \bibinfo
  {author} {\bibfnamefont {A.}~\bibnamefont {Hoffmann}}, \bibinfo {author}
  {\bibfnamefont {E.}~\bibnamefont {Gonzalez}}, \bibinfo {author}
  {\bibfnamefont {J.}~\bibnamefont {Vicent}}, \ and\ \bibinfo {author}
  {\bibfnamefont {I.~K.}\ \bibnamefont {Schuller}},\ }\href@noop {} {\bibfield
  {journal} {\bibinfo  {journal} {Journal of Magnetism and Magnetic Materials}\
  }\textbf {\bibinfo {volume} {320}},\ \bibinfo {pages} {2547} (\bibinfo {year}
  {2008})}\BibitemShut {NoStop}%
\bibitem [{\citenamefont {Moshchalkov}\ \emph {et~al.}(1998)\citenamefont
  {Moshchalkov}, \citenamefont {Baert}, \citenamefont {Metlushko},
  \citenamefont {Rosseel}, \citenamefont {Van~Bael}, \citenamefont {Temst},
  \citenamefont {Bruynseraede},\ and\ \citenamefont {Jonckheere}}]{Ref10}%
  \BibitemOpen
  \bibfield  {author} {\bibinfo {author} {\bibfnamefont {V.~V.}\ \bibnamefont
  {Moshchalkov}}, \bibinfo {author} {\bibfnamefont {M.}~\bibnamefont {Baert}},
  \bibinfo {author} {\bibfnamefont {V.~V.}\ \bibnamefont {Metlushko}}, \bibinfo
  {author} {\bibfnamefont {E.}~\bibnamefont {Rosseel}}, \bibinfo {author}
  {\bibfnamefont {M.~J.}\ \bibnamefont {Van~Bael}}, \bibinfo {author}
  {\bibfnamefont {K.}~\bibnamefont {Temst}}, \bibinfo {author} {\bibfnamefont
  {Y.}~\bibnamefont {Bruynseraede}}, \ and\ \bibinfo {author} {\bibfnamefont
  {R.}~\bibnamefont {Jonckheere}},\ }\href {\doibase 10.1103/PhysRevB.57.3615}
  {\bibfield  {journal} {\bibinfo  {journal} {Phys. Rev. B}\ }\textbf {\bibinfo
  {volume} {57}},\ \bibinfo {pages} {3615} (\bibinfo {year}
  {1998})}\BibitemShut {NoStop}%
\bibitem [{\citenamefont {Welp}\ \emph {et~al.}(2002)\citenamefont {Welp},
  \citenamefont {Xiao}, \citenamefont {Jiang}, \citenamefont {Vlasko-Vlasov},
  \citenamefont {Bader}, \citenamefont {Crabtree}, \citenamefont {Liang},
  \citenamefont {Chik},\ and\ \citenamefont {Xu}}]{Ref11}%
  \BibitemOpen
  \bibfield  {author} {\bibinfo {author} {\bibfnamefont {U.}~\bibnamefont
  {Welp}}, \bibinfo {author} {\bibfnamefont {Z.~L.}\ \bibnamefont {Xiao}},
  \bibinfo {author} {\bibfnamefont {J.~S.}\ \bibnamefont {Jiang}}, \bibinfo
  {author} {\bibfnamefont {V.~K.}\ \bibnamefont {Vlasko-Vlasov}}, \bibinfo
  {author} {\bibfnamefont {S.~D.}\ \bibnamefont {Bader}}, \bibinfo {author}
  {\bibfnamefont {G.~W.}\ \bibnamefont {Crabtree}}, \bibinfo {author}
  {\bibfnamefont {J.}~\bibnamefont {Liang}}, \bibinfo {author} {\bibfnamefont
  {H.}~\bibnamefont {Chik}}, \ and\ \bibinfo {author} {\bibfnamefont {J.~M.}\
  \bibnamefont {Xu}},\ }\href {\doibase 10.1103/PhysRevB.66.212507} {\bibfield
  {journal} {\bibinfo  {journal} {Phys. Rev. B}\ }\textbf {\bibinfo {volume}
  {66}},\ \bibinfo {pages} {212507} (\bibinfo {year} {2002})}\BibitemShut
  {NoStop}%
\bibitem [{\citenamefont {Silhanek}\ \emph {et~al.}(2005)\citenamefont
  {Silhanek}, \citenamefont {Van~Look}, \citenamefont {Jonckheere},
  \citenamefont {Zhu}, \citenamefont {Raedts},\ and\ \citenamefont
  {Moshchalkov}}]{Ref12}%
  \BibitemOpen
  \bibfield  {author} {\bibinfo {author} {\bibfnamefont {A.~V.}\ \bibnamefont
  {Silhanek}}, \bibinfo {author} {\bibfnamefont {L.}~\bibnamefont {Van~Look}},
  \bibinfo {author} {\bibfnamefont {R.}~\bibnamefont {Jonckheere}}, \bibinfo
  {author} {\bibfnamefont {B.~Y.}\ \bibnamefont {Zhu}}, \bibinfo {author}
  {\bibfnamefont {S.}~\bibnamefont {Raedts}}, \ and\ \bibinfo {author}
  {\bibfnamefont {V.~V.}\ \bibnamefont {Moshchalkov}},\ }\href {\doibase
  10.1103/PhysRevB.72.014507} {\bibfield  {journal} {\bibinfo  {journal} {Phys.
  Rev. B}\ }\textbf {\bibinfo {volume} {72}},\ \bibinfo {pages} {014507}
  (\bibinfo {year} {2005})}\BibitemShut {NoStop}%
\bibitem [{\citenamefont {Ooi}\ \emph {et~al.}(2009)\citenamefont {Ooi},
  \citenamefont {Mochiku},\ and\ \citenamefont {Hirata}}]{Ref13}%
  \BibitemOpen
  \bibfield  {author} {\bibinfo {author} {\bibfnamefont {S.}~\bibnamefont
  {Ooi}}, \bibinfo {author} {\bibfnamefont {T.}~\bibnamefont {Mochiku}}, \ and\
  \bibinfo {author} {\bibfnamefont {K.}~\bibnamefont {Hirata}},\ }\href@noop {}
  {\bibfield  {journal} {\bibinfo  {journal} {Physica C}\ }\textbf {\bibinfo
  {volume} {469}},\ \bibinfo {pages} {1113} (\bibinfo {year}
  {2009})}\BibitemShut {NoStop}%
\bibitem [{\citenamefont {Kemmler}\ \emph {et~al.}(2006)\citenamefont
  {Kemmler}, \citenamefont {G\"urlich}, \citenamefont {Sterck}, \citenamefont
  {P\"ohler}, \citenamefont {Neuhaus}, \citenamefont {Siegel}, \citenamefont
  {Kleiner},\ and\ \citenamefont {Koelle}}]{Ref14}%
  \BibitemOpen
  \bibfield  {author} {\bibinfo {author} {\bibfnamefont {M.}~\bibnamefont
  {Kemmler}}, \bibinfo {author} {\bibfnamefont {C.}~\bibnamefont {G\"urlich}},
  \bibinfo {author} {\bibfnamefont {A.}~\bibnamefont {Sterck}}, \bibinfo
  {author} {\bibfnamefont {H.}~\bibnamefont {P\"ohler}}, \bibinfo {author}
  {\bibfnamefont {M.}~\bibnamefont {Neuhaus}}, \bibinfo {author} {\bibfnamefont
  {M.}~\bibnamefont {Siegel}}, \bibinfo {author} {\bibfnamefont
  {R.}~\bibnamefont {Kleiner}}, \ and\ \bibinfo {author} {\bibfnamefont
  {D.}~\bibnamefont {Koelle}},\ }\href {\doibase 10.1103/PhysRevLett.97.147003}
  {\bibfield  {journal} {\bibinfo  {journal} {Phys. Rev. Lett.}\ }\textbf
  {\bibinfo {volume} {97}},\ \bibinfo {pages} {147003} (\bibinfo {year}
  {2006})}\BibitemShut {NoStop}%
\bibitem [{\citenamefont {Xu}\ \emph {et~al.}(2010)\citenamefont {Xu},
  \citenamefont {Cao},\ and\ \citenamefont {Heath}}]{Ref15}%
  \BibitemOpen
  \bibfield  {author} {\bibinfo {author} {\bibfnamefont {K.}~\bibnamefont
  {Xu}}, \bibinfo {author} {\bibfnamefont {P.}~\bibnamefont {Cao}}, \ and\
  \bibinfo {author} {\bibfnamefont {J.~R.}\ \bibnamefont {Heath}},\ }\bibfield
  {booktitle} {\emph {\bibinfo {booktitle} {Nano Letters}},\ }\href {\doibase
  10.1021/nl102584j} {\bibfield  {journal} {\bibinfo  {journal} {Nano Lett.}\
  }\textbf {\bibinfo {volume} {10}},\ \bibinfo {pages} {4206} (\bibinfo {year}
  {2010})}\BibitemShut {NoStop}%
\bibitem [{\citenamefont {Cao}\ \emph {et~al.}(2010)\citenamefont {Cao},
  \citenamefont {Horng}, \citenamefont {Wu}, \citenamefont {Yang},\ and\
  \citenamefont {Wu}}]{Ref16}%
  \BibitemOpen
  \bibfield  {author} {\bibinfo {author} {\bibfnamefont {R.}~\bibnamefont
  {Cao}}, \bibinfo {author} {\bibfnamefont {L.}~\bibnamefont {Horng}}, \bibinfo
  {author} {\bibfnamefont {J.}~\bibnamefont {Wu}}, \bibinfo {author}
  {\bibfnamefont {T.}~\bibnamefont {Yang}}, \ and\ \bibinfo {author}
  {\bibfnamefont {T.}~\bibnamefont {Wu}},\ }\href {\doibase
  10.1007/s10948-010-0729-5} {\bibfield  {journal} {\bibinfo  {journal}
  {Journal of Superconductivity and Novel Magnetism}\ }\textbf {\bibinfo
  {volume} {23}},\ \bibinfo {pages} {1051} (\bibinfo {year}
  {2010})}\BibitemShut {NoStop}%
\bibitem [{\citenamefont {Avci}\ \emph {et~al.}(2010)\citenamefont {Avci},
  \citenamefont {Xiao}, \citenamefont {Hua}, \citenamefont {Imre},
  \citenamefont {Divan}, \citenamefont {Pearson}, \citenamefont {Welp},
  \citenamefont {Kwok},\ and\ \citenamefont {Crabtree}}]{Ref17}%
  \BibitemOpen
  \bibfield  {author} {\bibinfo {author} {\bibfnamefont {S.}~\bibnamefont
  {Avci}}, \bibinfo {author} {\bibfnamefont {Z.~L.}\ \bibnamefont {Xiao}},
  \bibinfo {author} {\bibfnamefont {J.}~\bibnamefont {Hua}}, \bibinfo {author}
  {\bibfnamefont {A.}~\bibnamefont {Imre}}, \bibinfo {author} {\bibfnamefont
  {R.}~\bibnamefont {Divan}}, \bibinfo {author} {\bibfnamefont
  {J.}~\bibnamefont {Pearson}}, \bibinfo {author} {\bibfnamefont
  {U.}~\bibnamefont {Welp}}, \bibinfo {author} {\bibfnamefont {W.~K.}\
  \bibnamefont {Kwok}}, \ and\ \bibinfo {author} {\bibfnamefont {G.~W.}\
  \bibnamefont {Crabtree}},\ }\href {\doibase 10.1063/1.3473783} {\bibfield
  {journal} {\bibinfo  {journal} {Applied Physics Letters}\ }\textbf {\bibinfo
  {volume} {97}},\ \bibinfo {eid} {042511} (\bibinfo {year}
  {2010})}\BibitemShut {NoStop}%
\bibitem [{\citenamefont {Chiliotte}\ \emph {et~al.}(2011)\citenamefont
  {Chiliotte}, \citenamefont {Pasquini}, \citenamefont {Bekeris}, \citenamefont
  {Villegas}, \citenamefont {Li},\ and\ \citenamefont {Schuller}}]{Ref18}%
  \BibitemOpen
  \bibfield  {author} {\bibinfo {author} {\bibfnamefont {C.}~\bibnamefont
  {Chiliotte}}, \bibinfo {author} {\bibfnamefont {G.}~\bibnamefont {Pasquini}},
  \bibinfo {author} {\bibfnamefont {V.}~\bibnamefont {Bekeris}}, \bibinfo
  {author} {\bibfnamefont {J.~E.}\ \bibnamefont {Villegas}}, \bibinfo {author}
  {\bibfnamefont {C.-P.}\ \bibnamefont {Li}}, \ and\ \bibinfo {author}
  {\bibfnamefont {I.~K.}\ \bibnamefont {Schuller}},\ }\href@noop {} {\bibfield
  {journal} {\bibinfo  {journal} {Superconductor Science and Technology}\
  }\textbf {\bibinfo {volume} {24}},\ \bibinfo {pages} {065008} (\bibinfo
  {year} {2011})}\BibitemShut {NoStop}%
\bibitem [{\citenamefont {Zhang}\ \emph {et~al.}(2012)\citenamefont {Zhang},
  \citenamefont {He}, \citenamefont {Liu}, \citenamefont {Xue}, \citenamefont
  {Xiao}, \citenamefont {Li}, \citenamefont {Wen}, \citenamefont {Han},
  \citenamefont {Zhao}, \citenamefont {Gu}, \citenamefont {Qiu},\ and\
  \citenamefont {Moshchalkov}}]{Ref19}%
  \BibitemOpen
  \bibfield  {author} {\bibinfo {author} {\bibfnamefont {W.~J.}\ \bibnamefont
  {Zhang}}, \bibinfo {author} {\bibfnamefont {S.~K.}\ \bibnamefont {He}},
  \bibinfo {author} {\bibfnamefont {H.~F.}\ \bibnamefont {Liu}}, \bibinfo
  {author} {\bibfnamefont {G.~M.}\ \bibnamefont {Xue}}, \bibinfo {author}
  {\bibfnamefont {H.}~\bibnamefont {Xiao}}, \bibinfo {author} {\bibfnamefont
  {B.~H.}\ \bibnamefont {Li}}, \bibinfo {author} {\bibfnamefont {Z.~C.}\
  \bibnamefont {Wen}}, \bibinfo {author} {\bibfnamefont {X.~F.}\ \bibnamefont
  {Han}}, \bibinfo {author} {\bibfnamefont {S.~P.}\ \bibnamefont {Zhao}},
  \bibinfo {author} {\bibfnamefont {C.~Z.}\ \bibnamefont {Gu}}, \bibinfo
  {author} {\bibfnamefont {X.~G.}\ \bibnamefont {Qiu}}, \ and\ \bibinfo
  {author} {\bibfnamefont {V.~V.}\ \bibnamefont {Moshchalkov}},\ }\href@noop {}
  {\bibfield  {journal} {\bibinfo  {journal} {EPL (Europhysics Letters)}\
  }\textbf {\bibinfo {volume} {99}},\ \bibinfo {pages} {37006} (\bibinfo {year}
  {2012})}\BibitemShut {NoStop}%
\bibitem [{\citenamefont {Harada}\ \emph {et~al.}(1996)\citenamefont {Harada},
  \citenamefont {Kamimura}, \citenamefont {Kasai}, \citenamefont {Matsuda},
  \citenamefont {Tonomura},\ and\ \citenamefont {Moshchalkov}}]{New20}%
  \BibitemOpen
  \bibfield  {author} {\bibinfo {author} {\bibfnamefont {K.}~\bibnamefont
  {Harada}}, \bibinfo {author} {\bibfnamefont {O.}~\bibnamefont {Kamimura}},
  \bibinfo {author} {\bibfnamefont {H.}~\bibnamefont {Kasai}}, \bibinfo
  {author} {\bibfnamefont {T.}~\bibnamefont {Matsuda}}, \bibinfo {author}
  {\bibfnamefont {A.}~\bibnamefont {Tonomura}}, \ and\ \bibinfo {author}
  {\bibfnamefont {V.~V.}\ \bibnamefont {Moshchalkov}},\ }\href
  {http://www.sciencemag.org/content/274/5290/1167.abstract N2 - The
  microscopic mechanism of the matching effect in a superconductor, which
  manifested itself as the production of peaks or cusps in the critical current
  at specific values of the applied magnetic field, was investigated with
  Lorentz microscopy to allow direct observation of the behavior of vortices in
  a niobium thin film having a regular array of artificial defects. Vortices
  were observed to form regular and consequently rigid lattices at the matching
  magnetic field, at its multiples, and at its fractions. The dynamic
  observation furthermore revealed that vortices were most difficult to move at
  the matching field, whereas excess vortices moved easily.} {\bibfield
  {journal} {\bibinfo  {journal} {Science}\ }\textbf {\bibinfo {volume}
  {274}},\ \bibinfo {pages} {1167} (\bibinfo {year} {1996})}\BibitemShut
  {NoStop}%
\bibitem [{\citenamefont {Berdiyorov}\ \emph
  {et~al.}(2006{\natexlab{a}})\citenamefont {Berdiyorov}, \citenamefont
  {Milo\ifmmode \check{s}\else \v{s}\fi{}evi\ifmmode~\acute{c}\else
  \'{c}\fi{}},\ and\ \citenamefont {Peeters}}]{New21}%
  \BibitemOpen
  \bibfield  {author} {\bibinfo {author} {\bibfnamefont {G.~R.}\ \bibnamefont
  {Berdiyorov}}, \bibinfo {author} {\bibfnamefont {M.~V.}\ \bibnamefont
  {Milo\ifmmode \check{s}\else \v{s}\fi{}evi\ifmmode~\acute{c}\else
  \'{c}\fi{}}}, \ and\ \bibinfo {author} {\bibfnamefont {F.~M.}\ \bibnamefont
  {Peeters}},\ }\href {\doibase 10.1103/PhysRevLett.96.207001} {\bibfield
  {journal} {\bibinfo  {journal} {Phys. Rev. Lett.}\ }\textbf {\bibinfo
  {volume} {96}},\ \bibinfo {pages} {207001} (\bibinfo {year}
  {2006}{\natexlab{a}})}\BibitemShut {NoStop}%
\bibitem [{\citenamefont {Misko}\ \emph {et~al.}(2005)\citenamefont {Misko},
  \citenamefont {Savel'ev},\ and\ \citenamefont {Nori}}]{Ref20}%
  \BibitemOpen
  \bibfield  {author} {\bibinfo {author} {\bibfnamefont {V.}~\bibnamefont
  {Misko}}, \bibinfo {author} {\bibfnamefont {S.}~\bibnamefont {Savel'ev}}, \
  and\ \bibinfo {author} {\bibfnamefont {F.}~\bibnamefont {Nori}},\ }\href
  {\doibase 10.1103/PhysRevLett.95.177007} {\bibfield  {journal} {\bibinfo
  {journal} {Phys. Rev. Lett.}\ }\textbf {\bibinfo {volume} {95}},\ \bibinfo
  {pages} {177007} (\bibinfo {year} {2005})}\BibitemShut {NoStop}%
\bibitem [{\citenamefont {Misko}\ \emph {et~al.}(2006)\citenamefont {Misko},
  \citenamefont {Savel'ev},\ and\ \citenamefont {Nori}}]{Ref21}%
  \BibitemOpen
  \bibfield  {author} {\bibinfo {author} {\bibfnamefont {V.~R.}\ \bibnamefont
  {Misko}}, \bibinfo {author} {\bibfnamefont {S.}~\bibnamefont {Savel'ev}}, \
  and\ \bibinfo {author} {\bibfnamefont {F.}~\bibnamefont {Nori}},\ }\href
  {\doibase 10.1103/PhysRevB.74.024522} {\bibfield  {journal} {\bibinfo
  {journal} {Phys. Rev. B}\ }\textbf {\bibinfo {volume} {74}},\ \bibinfo
  {pages} {024522} (\bibinfo {year} {2006})}\BibitemShut {NoStop}%
\bibitem [{\citenamefont {Misko}\ \emph {et~al.}(2010)\citenamefont {Misko},
  \citenamefont {Bothner}, \citenamefont {Kemmler}, \citenamefont {Kleiner},
  \citenamefont {Koelle}, \citenamefont {Peeters},\ and\ \citenamefont
  {Nori}}]{Ref22}%
  \BibitemOpen
  \bibfield  {author} {\bibinfo {author} {\bibfnamefont {V.~R.}\ \bibnamefont
  {Misko}}, \bibinfo {author} {\bibfnamefont {D.}~\bibnamefont {Bothner}},
  \bibinfo {author} {\bibfnamefont {M.}~\bibnamefont {Kemmler}}, \bibinfo
  {author} {\bibfnamefont {R.}~\bibnamefont {Kleiner}}, \bibinfo {author}
  {\bibfnamefont {D.}~\bibnamefont {Koelle}}, \bibinfo {author} {\bibfnamefont
  {F.~M.}\ \bibnamefont {Peeters}}, \ and\ \bibinfo {author} {\bibfnamefont
  {F.}~\bibnamefont {Nori}},\ }\href {\doibase 10.1103/PhysRevB.82.184512}
  {\bibfield  {journal} {\bibinfo  {journal} {Phys. Rev. B}\ }\textbf {\bibinfo
  {volume} {82}},\ \bibinfo {pages} {184512} (\bibinfo {year}
  {2010})}\BibitemShut {NoStop}%
\bibitem [{\citenamefont {Kramer}\ \emph {et~al.}(2009)\citenamefont {Kramer},
  \citenamefont {Silhanek}, \citenamefont {Van~de Vondel}, \citenamefont
  {Raes},\ and\ \citenamefont {Moshchalkov}}]{Ref23}%
  \BibitemOpen
  \bibfield  {author} {\bibinfo {author} {\bibfnamefont {R.~B.~G.}\
  \bibnamefont {Kramer}}, \bibinfo {author} {\bibfnamefont {A.~V.}\
  \bibnamefont {Silhanek}}, \bibinfo {author} {\bibfnamefont {J.}~\bibnamefont
  {Van~de Vondel}}, \bibinfo {author} {\bibfnamefont {B.}~\bibnamefont {Raes}},
  \ and\ \bibinfo {author} {\bibfnamefont {V.~V.}\ \bibnamefont
  {Moshchalkov}},\ }\href {\doibase 10.1103/PhysRevLett.103.067007} {\bibfield
  {journal} {\bibinfo  {journal} {Phys. Rev. Lett.}\ }\textbf {\bibinfo
  {volume} {103}},\ \bibinfo {pages} {067007} (\bibinfo {year}
  {2009})}\BibitemShut {NoStop}%
\bibitem [{\citenamefont {Silhanek}\ \emph {et~al.}(2006)\citenamefont
  {Silhanek}, \citenamefont {Gillijns}, \citenamefont {Moshchalkov},
  \citenamefont {Zhu}, \citenamefont {Moonens},\ and\ \citenamefont
  {Leunissen}}]{Ref24}%
  \BibitemOpen
  \bibfield  {author} {\bibinfo {author} {\bibfnamefont {A.~V.}\ \bibnamefont
  {Silhanek}}, \bibinfo {author} {\bibfnamefont {W.}~\bibnamefont {Gillijns}},
  \bibinfo {author} {\bibfnamefont {V.~V.}\ \bibnamefont {Moshchalkov}},
  \bibinfo {author} {\bibfnamefont {B.~Y.}\ \bibnamefont {Zhu}}, \bibinfo
  {author} {\bibfnamefont {J.}~\bibnamefont {Moonens}}, \ and\ \bibinfo
  {author} {\bibfnamefont {L.~H.~A.}\ \bibnamefont {Leunissen}},\ }\href
  {\doibase 10.1063/1.2361172} {\bibfield  {journal} {\bibinfo  {journal}
  {Applied Physics Letters}\ }\textbf {\bibinfo {volume} {89}},\ \bibinfo {eid}
  {152507} (\bibinfo {year} {2006})}\BibitemShut {NoStop}%
\bibitem [{\citenamefont {Wu}\ \emph {et~al.}(2007)\citenamefont {Wu},
  \citenamefont {Horng}, \citenamefont {Wu}, \citenamefont {Cao}, \citenamefont
  {Kol\'{a}\v{c}ek},\ and\ \citenamefont {Yang}}]{Ref25}%
  \BibitemOpen
  \bibfield  {author} {\bibinfo {author} {\bibfnamefont {T.~C.}\ \bibnamefont
  {Wu}}, \bibinfo {author} {\bibfnamefont {L.}~\bibnamefont {Horng}}, \bibinfo
  {author} {\bibfnamefont {J.~C.}\ \bibnamefont {Wu}}, \bibinfo {author}
  {\bibfnamefont {R.}~\bibnamefont {Cao}}, \bibinfo {author} {\bibfnamefont
  {J.}~\bibnamefont {Kol\'{a}\v{c}ek}}, \ and\ \bibinfo {author} {\bibfnamefont
  {T.~J.}\ \bibnamefont {Yang}},\ }\href {\doibase 10.1063/1.2767386}
  {\bibfield  {journal} {\bibinfo  {journal} {Journal of Applied Physics}\
  }\textbf {\bibinfo {volume} {102}},\ \bibinfo {eid} {033918} (\bibinfo {year}
  {2007})}\BibitemShut {NoStop}%
\bibitem [{\citenamefont {Motta}\ \emph {et~al.}(2013)\citenamefont {Motta},
  \citenamefont {Colauto}, \citenamefont {Ortiz}, \citenamefont {Fritzsche},
  \citenamefont {Cuppens}, \citenamefont {Gillijns}, \citenamefont
  {Moshchalkov}, \citenamefont {Johansen}, \citenamefont {Sanchez},\ and\
  \citenamefont {Silhanek}}]{Ref26}%
  \BibitemOpen
  \bibfield  {author} {\bibinfo {author} {\bibfnamefont {M.}~\bibnamefont
  {Motta}}, \bibinfo {author} {\bibfnamefont {F.}~\bibnamefont {Colauto}},
  \bibinfo {author} {\bibfnamefont {W.~A.}\ \bibnamefont {Ortiz}}, \bibinfo
  {author} {\bibfnamefont {J.}~\bibnamefont {Fritzsche}}, \bibinfo {author}
  {\bibfnamefont {J.}~\bibnamefont {Cuppens}}, \bibinfo {author} {\bibfnamefont
  {W.}~\bibnamefont {Gillijns}}, \bibinfo {author} {\bibfnamefont {V.~V.}\
  \bibnamefont {Moshchalkov}}, \bibinfo {author} {\bibfnamefont {T.~H.}\
  \bibnamefont {Johansen}}, \bibinfo {author} {\bibfnamefont {A.}~\bibnamefont
  {Sanchez}}, \ and\ \bibinfo {author} {\bibfnamefont {A.~V.}\ \bibnamefont
  {Silhanek}},\ }\href@noop {} {\enquote {\bibinfo {title} {Enhancement of
  pinning properties of superconducting thin films by graded pinning
  landscapes.}}\ }\bibinfo {howpublished} {Preprint (arXiv: 1301.6283v1)}
  (\bibinfo {year} {2013})\BibitemShut {NoStop}%
\bibitem [{\citenamefont {Misko}\ and\ \citenamefont {Nori}(2012)}]{Ref27}%
  \BibitemOpen
  \bibfield  {author} {\bibinfo {author} {\bibfnamefont {V.~R.}\ \bibnamefont
  {Misko}}\ and\ \bibinfo {author} {\bibfnamefont {F.}~\bibnamefont {Nori}},\
  }\href {\doibase 10.1103/PhysRevB.85.184506} {\bibfield  {journal} {\bibinfo
  {journal} {Phys. Rev. B}\ }\textbf {\bibinfo {volume} {85}},\ \bibinfo
  {pages} {184506} (\bibinfo {year} {2012})}\BibitemShut {NoStop}%
\bibitem [{\citenamefont {Ray}\ \emph {et~al.}(2013)\citenamefont {Ray},
  \citenamefont {Olson~Reichhardt}, \citenamefont {Janko},\ and\ \citenamefont
  {Reichhardt}}]{Ref28}%
  \BibitemOpen
  \bibfield  {author} {\bibinfo {author} {\bibfnamefont {D.}~\bibnamefont
  {Ray}}, \bibinfo {author} {\bibfnamefont {C.~J.}\ \bibnamefont
  {Olson~Reichhardt}}, \bibinfo {author} {\bibfnamefont {B.}~\bibnamefont
  {Janko}}, \ and\ \bibinfo {author} {\bibfnamefont {C.}~\bibnamefont
  {Reichhardt}},\ }\href@noop {} {\enquote {\bibinfo {title} {Strongly enhanced
  vortex pinning by conformal crystal arrays.}}\ }\bibinfo {howpublished}
  {arXiv:1210.1229v1} (\bibinfo {year} {2013})\BibitemShut {NoStop}%
\bibitem [{\citenamefont {Rothen}\ \emph {et~al.}(1993)\citenamefont {Rothen},
  \citenamefont {Pieranski}, \citenamefont {Rivier},\ and\ \citenamefont
  {Joyet}}]{Ref29}%
  \BibitemOpen
  \bibfield  {author} {\bibinfo {author} {\bibfnamefont {F.}~\bibnamefont
  {Rothen}}, \bibinfo {author} {\bibfnamefont {P.}~\bibnamefont {Pieranski}},
  \bibinfo {author} {\bibfnamefont {N.}~\bibnamefont {Rivier}}, \ and\ \bibinfo
  {author} {\bibfnamefont {A.}~\bibnamefont {Joyet}},\ }\href@noop {}
  {\bibfield  {journal} {\bibinfo  {journal} {European Journal of Physics}\
  }\textbf {\bibinfo {volume} {14}},\ \bibinfo {pages} {227} (\bibinfo {year}
  {1993})}\BibitemShut {NoStop}%
\bibitem [{\citenamefont {Rothen}\ and\ \citenamefont
  {Piera\ifmmode~\acute{n}\else \'{n}\fi{}ski}(1996)}]{Ref30}%
  \BibitemOpen
  \bibfield  {author} {\bibinfo {author} {\bibfnamefont {F.~m.~c.}\
  \bibnamefont {Rothen}}\ and\ \bibinfo {author} {\bibfnamefont
  {P.}~\bibnamefont {Piera\ifmmode~\acute{n}\else \'{n}\fi{}ski}},\ }\href
  {\doibase 10.1103/PhysRevE.53.2828} {\bibfield  {journal} {\bibinfo
  {journal} {Phys. Rev. E}\ }\textbf {\bibinfo {volume} {53}},\ \bibinfo
  {pages} {2828} (\bibinfo {year} {1996})}\BibitemShut {NoStop}%
\bibitem [{\citenamefont {Yao}\ and\ \citenamefont {Jiang}(2011)}]{Ref31}%
  \BibitemOpen
  \bibfield  {author} {\bibinfo {author} {\bibfnamefont {K.}~\bibnamefont
  {Yao}}\ and\ \bibinfo {author} {\bibfnamefont {X.}~\bibnamefont {Jiang}},\
  }\href@noop {} {\bibfield  {journal} {\bibinfo  {journal} {J. Opt. Soc. Am.
  B}\ }\textbf {\bibinfo {volume} {28}},\ \bibinfo {pages} {1037} (\bibinfo
  {year} {2011})}\BibitemShut {NoStop}%
\bibitem [{\citenamefont {Garcia-Meca}\ \emph {et~al.}(2011)\citenamefont
  {Garcia-Meca}, \citenamefont {Martínez},\ and\ \citenamefont
  {Leonhardt}}]{Ref32}%
  \BibitemOpen
  \bibfield  {author} {\bibinfo {author} {\bibfnamefont {C.}~\bibnamefont
  {Garcia-Meca}}, \bibinfo {author} {\bibfnamefont {A.}~\bibnamefont
  {Martínez}}, \ and\ \bibinfo {author} {\bibfnamefont {U.}~\bibnamefont
  {Leonhardt}},\ }\href@noop {} {\bibfield  {journal} {\bibinfo  {journal}
  {Opt. Express}\ }\textbf {\bibinfo {volume} {19}},\ \bibinfo {pages} {23743}
  (\bibinfo {year} {2011})}\BibitemShut {NoStop}%
\bibitem [{\citenamefont {Baz\'{a}n}\ \emph {et~al.}(2007)\citenamefont
  {Baz\'{a}n}, \citenamefont {Torres}, \citenamefont {de~Espinosa},
  \citenamefont {Quintero-Torres},\ and\ \citenamefont {Arag\'{o}n}}]{Ref33}%
  \BibitemOpen
  \bibfield  {author} {\bibinfo {author} {\bibfnamefont {A.}~\bibnamefont
  {Baz\'{a}n}}, \bibinfo {author} {\bibfnamefont {M.}~\bibnamefont {Torres}},
  \bibinfo {author} {\bibfnamefont {F.~R.~M.}\ \bibnamefont {de~Espinosa}},
  \bibinfo {author} {\bibfnamefont {R.}~\bibnamefont {Quintero-Torres}}, \ and\
  \bibinfo {author} {\bibfnamefont {J.~L.}\ \bibnamefont {Arag\'{o}n}},\ }\href
  {\doibase 10.1063/1.2709939} {\bibfield  {journal} {\bibinfo  {journal}
  {Applied Physics Letters}\ }\textbf {\bibinfo {volume} {90}},\ \bibinfo {eid}
  {094101} (\bibinfo {year} {2007})}\BibitemShut {NoStop}%
\bibitem [{\citenamefont {Gu}\ \emph {et~al.}(2012)\citenamefont {Gu},
  \citenamefont {Yao}, \citenamefont {Lu}, \citenamefont {Lai}, \citenamefont
  {Chen}, \citenamefont {Hou},\ and\ \citenamefont {Jiang}}]{Ref34}%
  \BibitemOpen
  \bibfield  {author} {\bibinfo {author} {\bibfnamefont {C.}~\bibnamefont
  {Gu}}, \bibinfo {author} {\bibfnamefont {K.}~\bibnamefont {Yao}}, \bibinfo
  {author} {\bibfnamefont {W.}~\bibnamefont {Lu}}, \bibinfo {author}
  {\bibfnamefont {Y.}~\bibnamefont {Lai}}, \bibinfo {author} {\bibfnamefont
  {H.}~\bibnamefont {Chen}}, \bibinfo {author} {\bibfnamefont {B.}~\bibnamefont
  {Hou}}, \ and\ \bibinfo {author} {\bibfnamefont {X.}~\bibnamefont {Jiang}},\
  }\href {\doibase 10.1063/1.4731877} {\bibfield  {journal} {\bibinfo
  {journal} {Applied Physics Letters}\ }\textbf {\bibinfo {volume} {100}},\
  \bibinfo {eid} {261907} (\bibinfo {year} {2012})}\BibitemShut {NoStop}%
\bibitem [{\citenamefont {Liang}\ \emph {et~al.}(2010)\citenamefont {Liang},
  \citenamefont {Kunchur}, \citenamefont {Hua},\ and\ \citenamefont
  {Xiao}}]{Ref35}%
  \BibitemOpen
  \bibfield  {author} {\bibinfo {author} {\bibfnamefont {M.}~\bibnamefont
  {Liang}}, \bibinfo {author} {\bibfnamefont {M.~N.}\ \bibnamefont {Kunchur}},
  \bibinfo {author} {\bibfnamefont {J.}~\bibnamefont {Hua}}, \ and\ \bibinfo
  {author} {\bibfnamefont {Z.}~\bibnamefont {Xiao}},\ }\href {\doibase
  10.1103/PhysRevB.82.064502} {\bibfield  {journal} {\bibinfo  {journal} {Phys.
  Rev. B}\ }\textbf {\bibinfo {volume} {82}},\ \bibinfo {pages} {064502}
  (\bibinfo {year} {2010})}\BibitemShut {NoStop}%
\bibitem [{\citenamefont {Patel}\ \emph {et~al.}(2007)\citenamefont {Patel},
  \citenamefont {Xiao}, \citenamefont {Hua}, \citenamefont {Xu}, \citenamefont
  {Rosenmann}, \citenamefont {Novosad}, \citenamefont {Pearson}, \citenamefont
  {Welp}, \citenamefont {Kwok},\ and\ \citenamefont {Crabtree}}]{Ref36}%
  \BibitemOpen
  \bibfield  {author} {\bibinfo {author} {\bibfnamefont {U.}~\bibnamefont
  {Patel}}, \bibinfo {author} {\bibfnamefont {Z.~L.}\ \bibnamefont {Xiao}},
  \bibinfo {author} {\bibfnamefont {J.}~\bibnamefont {Hua}}, \bibinfo {author}
  {\bibfnamefont {T.}~\bibnamefont {Xu}}, \bibinfo {author} {\bibfnamefont
  {D.}~\bibnamefont {Rosenmann}}, \bibinfo {author} {\bibfnamefont
  {V.}~\bibnamefont {Novosad}}, \bibinfo {author} {\bibfnamefont
  {J.}~\bibnamefont {Pearson}}, \bibinfo {author} {\bibfnamefont
  {U.}~\bibnamefont {Welp}}, \bibinfo {author} {\bibfnamefont {W.~K.}\
  \bibnamefont {Kwok}}, \ and\ \bibinfo {author} {\bibfnamefont {G.~W.}\
  \bibnamefont {Crabtree}},\ }\href {\doibase 10.1103/PhysRevB.76.020508}
  {\bibfield  {journal} {\bibinfo  {journal} {Phys. Rev. B}\ }\textbf {\bibinfo
  {volume} {76}},\ \bibinfo {pages} {020508} (\bibinfo {year}
  {2007})}\BibitemShut {NoStop}%
\bibitem [{\citenamefont {Sochnikov}\ \emph {et~al.}(2010)\citenamefont
  {Sochnikov}, \citenamefont {Shaulov}, \citenamefont {Yeshurun}, \citenamefont
  {Logvenov},\ and\ \citenamefont {Bozovic}}]{New39}%
  \BibitemOpen
  \bibfield  {author} {\bibinfo {author} {\bibfnamefont {I.}~\bibnamefont
  {Sochnikov}}, \bibinfo {author} {\bibfnamefont {A.}~\bibnamefont {Shaulov}},
  \bibinfo {author} {\bibfnamefont {Y.}~\bibnamefont {Yeshurun}}, \bibinfo
  {author} {\bibfnamefont {G.}~\bibnamefont {Logvenov}}, \ and\ \bibinfo
  {author} {\bibfnamefont {I.}~\bibnamefont {Bozovic}},\ }\href
  {http://dx.doi.org/10.1038/nnano.2010.111} {\bibfield  {journal} {\bibinfo
  {journal} {Nat Nano}\ }\textbf {\bibinfo {volume} {5}},\ \bibinfo {pages}
  {516} (\bibinfo {year} {2010})}\BibitemShut {NoStop}%
\bibitem [{\citenamefont {Berdiyorov}\ \emph {et~al.}(2012)\citenamefont
  {Berdiyorov}, \citenamefont {Milo\ifmmode \check{s}\else
  \v{s}\fi{}evi\ifmmode~\acute{c}\else \'{c}\fi{}}, \citenamefont {Latimer},
  \citenamefont {Xiao}, \citenamefont {Kwok},\ and\ \citenamefont
  {Peeters}}]{New40}%
  \BibitemOpen
  \bibfield  {author} {\bibinfo {author} {\bibfnamefont {G.~R.}\ \bibnamefont
  {Berdiyorov}}, \bibinfo {author} {\bibfnamefont {M.~V.}\ \bibnamefont
  {Milo\ifmmode \check{s}\else \v{s}\fi{}evi\ifmmode~\acute{c}\else
  \'{c}\fi{}}}, \bibinfo {author} {\bibfnamefont {M.~L.}\ \bibnamefont
  {Latimer}}, \bibinfo {author} {\bibfnamefont {Z.~L.}\ \bibnamefont {Xiao}},
  \bibinfo {author} {\bibfnamefont {W.~K.}\ \bibnamefont {Kwok}}, \ and\
  \bibinfo {author} {\bibfnamefont {F.~M.}\ \bibnamefont {Peeters}},\ }\href
  {\doibase 10.1103/PhysRevLett.109.057004} {\bibfield  {journal} {\bibinfo
  {journal} {Phys. Rev. Lett.}\ }\textbf {\bibinfo {volume} {109}},\ \bibinfo
  {pages} {057004} (\bibinfo {year} {2012})}\BibitemShut {NoStop}%
\bibitem [{Sup()}]{Sup}%
  \BibitemOpen
  \href@noop {} {}\bibinfo {howpublished} {See Supplemental Material at [URL
  will be inserted by publisher] for sample fabrications and
  measurements.}\BibitemShut {Stop}%
\bibitem [{\citenamefont {Reichhardt}\ and\ \citenamefont
  {Olson~Reichhardt}(2009)}]{Ref37}%
  \BibitemOpen
  \bibfield  {author} {\bibinfo {author} {\bibfnamefont {C.}~\bibnamefont
  {Reichhardt}}\ and\ \bibinfo {author} {\bibfnamefont {C.~J.}\ \bibnamefont
  {Olson~Reichhardt}},\ }\href {\doibase 10.1103/PhysRevB.79.134501} {\bibfield
   {journal} {\bibinfo  {journal} {Phys. Rev. B}\ }\textbf {\bibinfo {volume}
  {79}},\ \bibinfo {pages} {134501} (\bibinfo {year} {2009})}\BibitemShut
  {NoStop}%
\bibitem [{\citenamefont {Latimer}\ \emph {et~al.}(2012)\citenamefont
  {Latimer}, \citenamefont {Berdiyorov}, \citenamefont {Xiao}, \citenamefont
  {Kwok},\ and\ \citenamefont {Peeters}}]{Ref38}%
  \BibitemOpen
  \bibfield  {author} {\bibinfo {author} {\bibfnamefont {M.~L.}\ \bibnamefont
  {Latimer}}, \bibinfo {author} {\bibfnamefont {G.~R.}\ \bibnamefont
  {Berdiyorov}}, \bibinfo {author} {\bibfnamefont {Z.~L.}\ \bibnamefont
  {Xiao}}, \bibinfo {author} {\bibfnamefont {W.~K.}\ \bibnamefont {Kwok}}, \
  and\ \bibinfo {author} {\bibfnamefont {F.~M.}\ \bibnamefont {Peeters}},\
  }\href {\doibase 10.1103/PhysRevB.85.012505} {\bibfield  {journal} {\bibinfo
  {journal} {Phys. Rev. B}\ }\textbf {\bibinfo {volume} {85}},\ \bibinfo
  {pages} {012505} (\bibinfo {year} {2012})}\BibitemShut {NoStop}%
\bibitem [{\citenamefont {Berdiyorov}\ \emph
  {et~al.}(2006{\natexlab{b}})\citenamefont {Berdiyorov}, \citenamefont
  {Milosevic},\ and\ \citenamefont {Peeters}}]{New44}%
  \BibitemOpen
  \bibfield  {author} {\bibinfo {author} {\bibfnamefont {G.~R.}\ \bibnamefont
  {Berdiyorov}}, \bibinfo {author} {\bibfnamefont {M.~V.}\ \bibnamefont
  {Milosevic}}, \ and\ \bibinfo {author} {\bibfnamefont {F.~M.}\ \bibnamefont
  {Peeters}},\ }\href {http://stacks.iop.org/0295-5075/74/i=3/a=493} {\bibfield
   {journal} {\bibinfo  {journal} {EPL (Europhysics Letters)}\ }\textbf
  {\bibinfo {volume} {74}},\ \bibinfo {pages} {493} (\bibinfo {year}
  {2006}{\natexlab{b}})}\BibitemShut {NoStop}%
\end{thebibliography}%

\end{document}